\documentclass[aps,twocolumn,superscriptaddress,footinbib,floatfix,showpacs]{revtex4-1} 
\usepackage{amsfonts}
\usepackage{amssymb}
\usepackage[fleqn]{amsmath}
\usepackage[mathscr]{eucal}
\usepackage{graphicx} 
\usepackage{subfigure}
\usepackage{comment,color}
\usepackage{natbib}
\def\r{\mathbf{r}}

\begin{document} 
  
\title{Ergodicity and spectral cascades in point vortex flows on the sphere}
      
\author{David G. Dritschel}
\affiliation{Mathematical Institute, University of St Andrews,
St Andrews,  KY16 9SS, UK}
\author{Marcello Lucia}
\author{Andrew C. Poje}
\affiliation{Graduate Faculty in Physics \& Department of Mathematics,
City University of New York - CSI, Staten Island, New York 10314} 

\begin{abstract} {We present results for the equilibrium statistics and 
     dynamic evolution of moderately large ($n =
    {\mathcal{O}}(10^2 - 10^3)$) numbers of interacting point vortices
    on the sphere under the constraint of zero mean angular
    momentum. For systems with equal numbers of
    positive and negative identical circulations,
    the density of re-scaled energies,
    $p(E)$, converges rapidly with $n$ to a
    function with a single maximum with maximum entropy.
    Ensemble-averaged wavenumber spectra of the nonsingular velocity field induced by
    the vortices
    exhibit the expected $k^{-1}$ behavior at
    small scales for all energies.  Spectra at the largest scales
    vary continuously with the inverse temperature of the system.
    For positive temperatures, spectra peak at finite 
    intermediate wavenumbers; 
    for negative temperatures, spectra decrease everywhere. 
    Comparisons of time and ensemble averages, over a large range of energies, 
    strongly support ergodicity in the dynamics even for highly atypical initial vortex 
    configurations. 
    Crucially, rapid relaxation of spectra towards the microcanonical average implies that the direction
    of any spectral cascade process depends only on the relative difference
    between the initial spectrum and the ensemble mean spectrum at
    that energy; not on the energy, or temperature, of the system.}
\end{abstract}
\maketitle

\section{Introduction}

The point vortex model, originally developed by Kirchoff 
\cite{Kirchhoff:1876} as a limiting form of Euler's equations in 
two-dimensions, continues to provide a conceptual and computational tool for understanding 
inviscid, nonlinear vortex dynamics in both traditional and superfluid turbulence (\cite{PhysRevE.89.013009, Wang2007, simula}).
The multi-body Hamiltonian describing the dynamics of idealized point vortices serves as a paradigm for
developing kinetic theories in systems dominated by long-range interactions 
\cite{Kiessling97, Chavanis12}.

The statistical mechanics of point vortex systems was first addressed in the seminal work of Onsager 
\cite{Onsager} who observed that in a finite domain, the Hamiltonian structure of a system of sufficiently 
large numbers of positive and negative vortices implies the existence of `negative temperature'
equilibrium states which naturally exhibit clustering of like-signed vortices \cite{EyinkSree:2006}.
Onsager's statistical approach has inspired a wealth 
of subsequent work on vortex-based, mean-field turbulence closures \cite{MontgomeryJoyce, *LundPoint, *KraichnanMontgomery,
*Miller, *Robert}
and the existence of negative temperature states has been interpreted 
(e.g. \cite{Buhler, yatsu, Wang2007, simula}) as an energy-conserving analog of self-organization via `vortex merger' 
commonly observed in two-dimensional turbulence \cite{Mac83, d08}. 
To date, however, direct connections between Onsager's equilibrium prediction for the inviscid point-vortex
system and the up-scaling, inverse-energy cascade in two-dimensional Navier-Stokes turbulence have proved elusive.

Underpinning the equilibrium statistical mechanics approach are the assumptions that the system
is both energy isolated (inviscid) and ergodic, namely that as $t \to \infty$, the system samples all possible configurations 
on a fixed energy surface.  While the inviscid assumption is clearly violated by Navier-Stokes vortices, 
two-dimensional turbulence cascades energy to the largest scales where viscous effects are less pronounced. 
In addition, as long as the relaxation to equilibrium of the inviscid system takes place on 
time-scales much shorter than those imposed by viscosity, the equilibrium statistics of the inviscid model should 
approximate those of the full system on these timescales \cite{EyinkSpohn:1993}. 
Ergodicity of point-vortex systems remains an open issue. The assumption was questioned by Onsager \cite{Onsager} and
repeatedly since \cite{wm91,Tabeling2002}. 

In the present work, we directly examine both ergodicity and the connection between equilibrium statistical properties and
dynamic kinetic energy cascades for two-dimensional point-vortex systems on the unit sphere
\cite{zermelo}. The sphere has the distinct advantage of providing a bounded domain without
the complications of imposing explicit boundary conditions via image particles (infinitely many
for doubly-periodic domains).
Despite its apparent attraction, there has been relatively little work addressing the statistical mechanics of point 
vortices on the sphere.  Recently, for spherical systems with skewed distributions of vortex strengths, 
Kiessling \& Wang (2012) \cite{kiessling2012}
proved convergence to continuous solutions of Euler's equations.
The scaling limits considered, however, assume the existence of
large-scale mean flows and thus have singular structure in the zero mean, zero angular
momentum limit.  

In closer analogy with turbulence studies, we study fluctuations in zero angular momentum states 
of binary populations of vortices with zero mean circulation (see, for example, \cite{Buhler,yatsu,wm91}). 
We find that the kinetic energy spectrum of flows induced by such systems 
scales as $k^{-1}$, for sufficiently large degree (or wavenumber)
$k$, independent of the system energy.
As Onsager conjectured, increasing the energy of the system necessarily increases the
kinetic energy content at the largest allowable scales.
However, comparisons of microcanonical and time averaged two-point statistics show clear evidence of ergodicity in the
vortex dynamics implying that the direction of any dynamic spectral evolution depends solely on the shape of the initial spectrum relative
to the ensemble mean. Therefore, for the spherical system, there is no a priori association between negative temperature states and 
the inverse energy cascade.

\section{Equilibrium Statistics}

Point vortices on a unit sphere evolve according to Hamilton's equations, 
with conserved Hamiltonian
\begin{equation} \label{eq:HamSphere}
   H = - \sum_{i=1}^n \sum_{j \neq i} \kappa_i \kappa_j 
             \ln \left[\left(  1 -  \mathbf{r}_i \cdot \mathbf{r}_j \right)/2 \right].
\end{equation}
Here $\kappa_i$ is the 
`strength' (circulation$/4\pi$) 
of vortex $i$ and $\r_i$ its
position ($|\r_i|=1$).  The evolution equations are
\begin{equation} \label{eq:Dynamics}
   \frac{d \r_i}{dt} = 2\sum_{j \neq i}
  \kappa_j \frac{ \r_i  \times \r_j}{1 -   \r_i \cdot \r_j}  .
\end{equation}
In addition to $H$, 
the vector 
angular impulse,
${\mathbf I} = \sum_{i=1}^n  \kappa_i \r_i$,
is also conserved although only the angular momentum,  $|{\mathbf{I}}|$, affects
the statistical properties.

We consider systems with $\kappa_i = \pm 1$, and zero net circulation.
The pairwise interaction energies are
\begin{equation} \label{eq:IJInteraction}
      q_{ij} = \pm \ln \left[ \left( 1 -  \r_i \cdot \r_j \right)/2  \right].
\end{equation}
For randomly placed vortices, the argument of the logarithm is
uniformly-distributed over $(0,1)$. Thus,
$q_{ij}$ is exponentially-distributed over $(0, \infty)$ where $\langle q_{ij} \rangle = 1$
and over $(-\infty, 0)$ where $\langle q_{ij} \rangle = - 1$.
In particular,
\begin{equation}
    \langle H \rangle
    =   \sum_{i=1}^n \sum_{j \neq i} \langle q_{ij} \rangle
    =   2  \left( \frac{n}{2} ( \frac{n}{2} -1) -  (\frac{n}{2})^2  \right)
     = - n .
\nonumber
\end{equation}
For any distribution of vortex strengths with identical numbers of opposite-signed circulations,  similar
cancellations occur and  
$\langle H \rangle = O(n)$ \cite{Buhler, yatsu,Esler} rather than  $\langle H \rangle = O(n^2)$
\cite{kiessling2012}.  Given exponential $q$ statistics, the standard deviation of $H$ 
is also $O(n)$.
In this case, the joint density of states, 
$W_{H}(\tilde{E},\tilde{J}) = $ 
\[
\int_{S^{2n}} \delta(\tilde{E} - H(\r_1,\ldots,\r_n)) \, \delta (\tilde{J}-|\mathbf{I}(\r_1,\ldots,\r_n)|) d\r_1 \ldots d\r_n
\]
has a limiting function 
$p(E,J)=\lim_{n \to \infty} n W_{H/n}(\tilde{E},\tilde{J})$ 
for the 
specific energy
$E = \tilde{E}/n$ and re-scaled angular momentum $J = \tilde{J}/ \sqrt{n}$.

The re-scaled density has been computed numerically by sampling
$10^9$ uniformly-distributed placements of $n=200$ vortices.
In this case, $\langle E \rangle=-1.0000$, as expected, with $\langle J \rangle= 0.9215$.  The observed distribution is asymmetric with a single 
maximum
at 
$\left(E,J \right) = \left(-1.684,0.824 \right)$, significantly different from the mean.

\begin{figure}[!htb]
\begin{center}
{\includegraphics[width=0.40\textwidth]{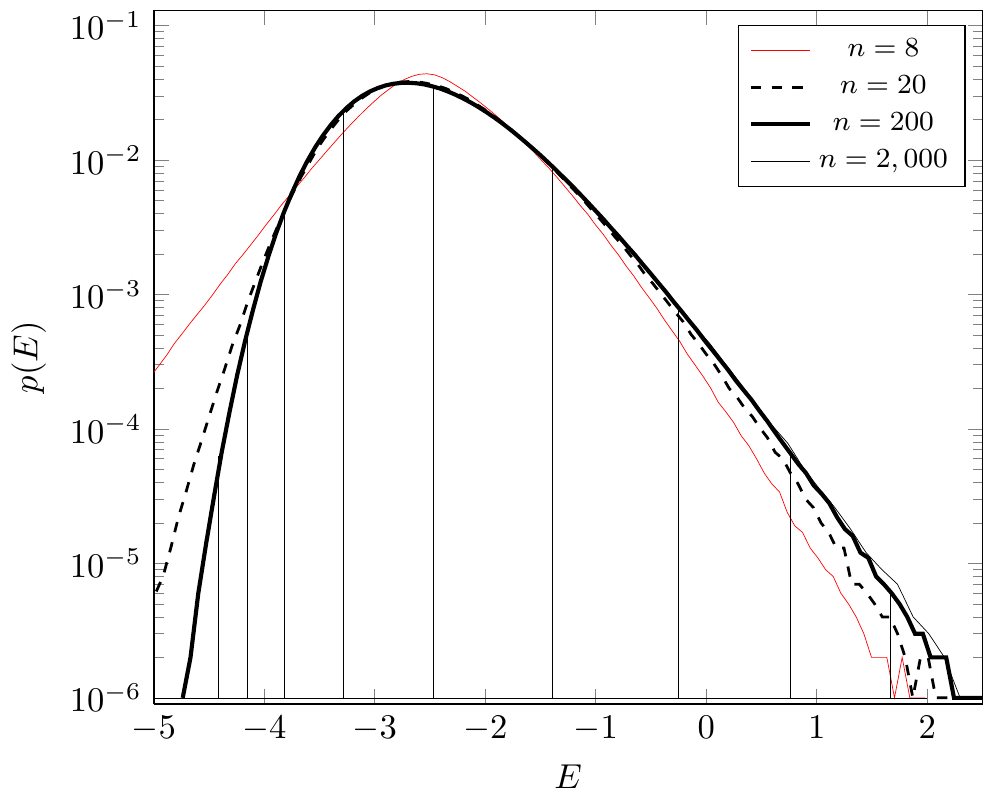}}
\put(-220,165){(a)}
\\
{\includegraphics[width=0.40\textwidth]{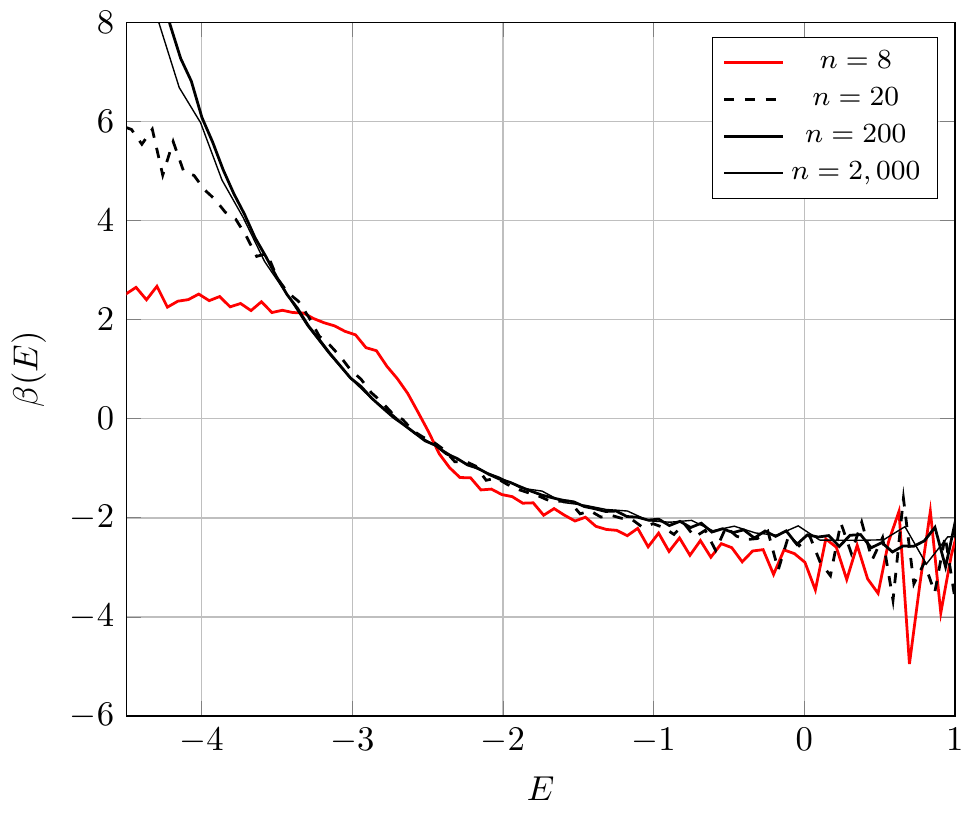}}
\put(-220,165){(b)}
\caption{(a) Distribution function $p(E)$ computed from $10^7$
  samples for different numbers of vortices $n$. Vertical lines
  correspond to the 9 energy levels for $n=200$ considered in the
  text.  (b) Corresponding inverse temperatures $\beta(E)$.}
\label{fig1}
\end{center}
\end{figure}
Direct extraction of $p(E) := p(E,J=0)$ from the joint density is 
computationally expensive; estimates can be obtained more efficiently by 
adjusting random states towards $J=0$.  From a single realization of $n$
randomly-generated vortex positions, we compute ${\mathbf I}$ and then
displace each vortex by $-\kappa_i{\mathbf I}/n$.  This sets $J=0$,
but the vortices no longer reside on the spherical surface.  Re-scaling
each $\r_i$ by $|\r_i|$ produces a new ${\mathbf I}$, and the process is 
iterated until convergence.  
For $n=200$, $p(E)$ computed this way was
found to be identical within sampling errors to 
$p(E,J<0.2)$ estimated from the joint density.

For fixed $n$, $p(E)$ was estimated by binning $10^7$ 
samples of $n$ uniformly distributed vortex positions iterated to $J < 10^{-14}$.
The resulting density and inverse temperature,
$\beta = d \ln{p(E)}/dE$,  are shown for varying $n$ in Fig.~\ref{fig1}.
While nearly symmetric for small $n$, the scaled density converges rapidly to a
skewed distribution as $n$ increases. The scaled inverse temperature asymptotes 
to a fixed, negative value at large positive energies \cite{PointLund, yatsu, Esler}.
There is little difference in either the density of states or the temperature 
when $n$ increases beyond 200. 

\section{Kinetic Energy Spectra} 
Much has been intimated about Onsager's statistical theory of 
self-organization and the widely-observed scale cascade of kinetic energy 
in direct simulations of two-dimensional turbulence 
\cite{EyinkSree:2006,Buhler,yatsu}.  The scale cascade results in the
accumulation of energy at the domain scale, i.e.\ a global-scale flow
\cite{qi14b}.

To compare the dynamic evolution 
of point vortices
to microcanonical ensemble
predictions, we consider two statistical measures
of the vortex population.  Both quantify any scale cascade or statistical
change in the vortex population, though neither have been examined before
in this context.
First, 
as in nearly all studies of two-dimensional turbulence,
%
we examine 
the kinetic energy
spectrum $K(k)$ where $k$ is the wavenumber magnitude 
(spherical harmonic degree).
$K(k)$
is calculated by evaluating the streamfunction
\begin{equation}
\psi(\r)=  \sum_{i=1}^n \kappa_i \ln \left[ \left( 1 -  \r_i \cdot \r \right)/2  \right]
\end{equation}
induced by the vortices at every point $\r$ on a regular
latitude-longitude grid (1024 $\times$ 2048 points).  The
Fourier-Legendre transform of $\psi$ and its 
(power)
spectrum 
$P(k)$
are then computed and we obtain $K(k)$ from $k(k+1)P(k)$.
While the total kinetic energy is singular as a result
of the $k^{-1}$ spectral tail, the spectrum $K(k)$ is well
behaved for finite $k$.

A complementary Lagrangian measure 
of the vortex population
is given by the probability
distribution $p_{\rm int}(q)$ of the pair-wise energy ~\eqref{eq:IJInteraction}.
To explicitly highlight anomalous distributions of dipoles or
like-signed clusters, we consider the residual probability 
$p'_{\rm int} \equiv p_{\rm int}-e^{-|q|}/2$
by subtracting the exponential distribution produced by uniform, random placement.

\begin{figure}[!htb]
\begin{center}
    \includegraphics[width=0.49\textwidth]{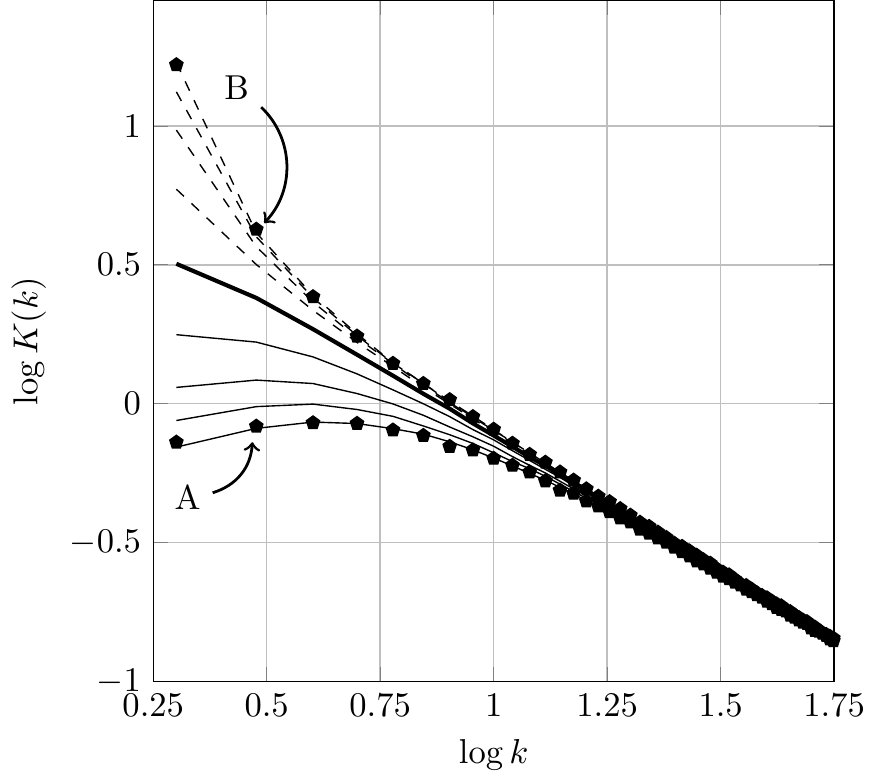}
\end{center}
\vspace{-0.25in}
\caption{Microcanonical kinetic energy spectra, $K(k)$ for the nine
  energies considered.  $K$ at low wavenumbers increases monotonically
  with energy $E$ from A to B. $\beta>0$ states shown in solid, $\beta <
  0$ states dashed and $\beta \sim 0$ in bold. Solid circles indicate time averages of dynamical evolution.}
\label{fig2}
\end{figure}
For $n=200$, these two statistics are computed by sampling $10^4$ states 
within each of
nine energy ranges 
centered around the vertical lines shown in Fig.~\ref{fig1}a. 
The energy ranges include both positive and negative temperature states,
and are narrow -- the probability of finding a state in a given range never exceeds $3.7 \times 10^{-5}$. 

All nine individual kinetic energy spectra shown in the upper panel of 
Fig.~\ref{fig2} 
converge to the expected $k^{-1}$ form at small scales.  Consistent with
Onsager's predictions, positive temperature (strongly negative $E$)
states have the least kinetic energy at largest scales.  The kinetic
energy content at the largest scales increases continuously as $E$
increases and the system transitions to negative temperature states.
Notably, the spectral slope at small $k$ changes from values above $-1$
to below $-1$ near $\beta = 0$.

\begin{figure}[!htb]
\begin{center}
    \includegraphics[width=0.49\textwidth]{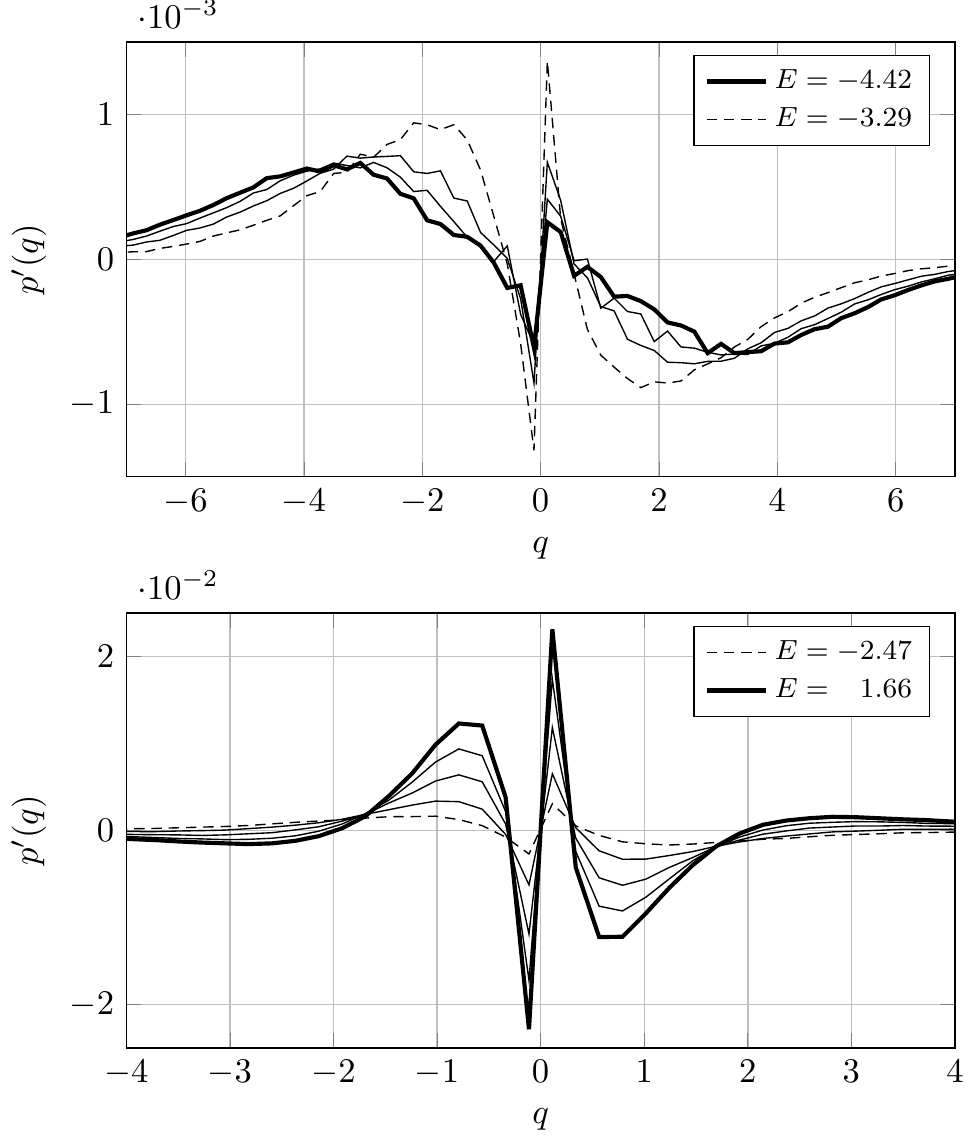}
\end{center}
\vspace{-0.25in}
\caption{ The residual probability %
  $p'_{\rm int}$
  versus normalized vortex interaction energy, $q$ for (a) 
  lower range of energies considered and (b) higher range of
  energies (note change of scales).} 
\label{fig3}
\end{figure}
 
The low energy ($\beta >0$) spectra are consistent with
\textit{dipole} spectra produced by randomly placing pairs of
opposite-signed vortices.  Such spectra are depleted at low $k$ and,
as $E$ decreases, approach $k^1$ at the large scales.  
The surplus of
dipoles for positive $\beta$ states is seen in $p'_{\rm int}(q)$ shown in
Fig.~\ref{fig3}.  
Like the kinetic energy spectrum, 
$p'_{\rm int}$ exhibits a monotonic dependence on $E$ with 
a surplus of closely-spaced dipoles having $q \ll -1$
at low $E$,
while at high $E$ ($\beta <0$) there is a surplus of
closely-spaced like-signed pairs (binaries) 
having $q \gg 1$
together with a deficit of
closely-spaced dipoles.  
Importantly, both complementary 
statistics, $\langle K \rangle(k)$ and $\langle p'_{\rm int} \rangle(q)$,
vary continuously with the inverse system temperature $\beta$.
There is no abrupt change in either at the transition from positive to
negative temperatures.

\section{Ergodicity and Spectral Cascade }
We now turn our attention to the question of ergodicity
by quantifying the connection between time-averaged statistics of dynamically evolved
states and microcanonical ensemble measures. 
The evolution equation (\ref{eq:Dynamics}) is solved in parallel 
using a 4th order Runge-Kutta scheme with an
adaptive time step to ensure exact conservation of momentum and energy preservation to $10^{-7}$.
As such, numerical variations in the dynamically evolved energy are always smaller than the
width of the energy bins used to construct microcanonical statistics.
With $n=200$, 
a single state in each of the 9 energy ranges 
was evolved for $400$ time units.
Redefining the vortex strengths as $\kappa_i=\pm 1/\sqrt{n}$,
gives $E=H$ directly from (\ref{eq:HamSphere}) and a characteristic timescale is then
$\tau=\pi\bar{d}^2/|\kappa_i|=4\pi^2/\sqrt{n}$, approximately
$2.79$ for $n=200$. 

The kinetic energy spectra and $\langle p'_{\rm int}
\rangle(q)$,  time-averaged over the entire evolution
were found to be almost identical to the
microcanonical ensemble results.  The resulting time averaged kinetic energy spectra, 
$\overline{K}(k)$, for the two extreme energies $E =
-4.42$ and $1.66$ indicated by solid circles in Fig.~\ref{fig2} are
virtually indistinguishable from the microcanonical estimates. 
The same is found for $\langle p'_{\rm int}
\rangle(q)$. 
In contrast to previous results for $n = 6$ vortices in a doubly-periodic domain \cite{wm91},  
here for $n=200$ vortices on the sphere there is strong evidence of 
ergodicity, independent of the energy or temperature of the system.

\begin{figure}[!htb]
\begin{center}
    \includegraphics[width=0.49\textwidth]{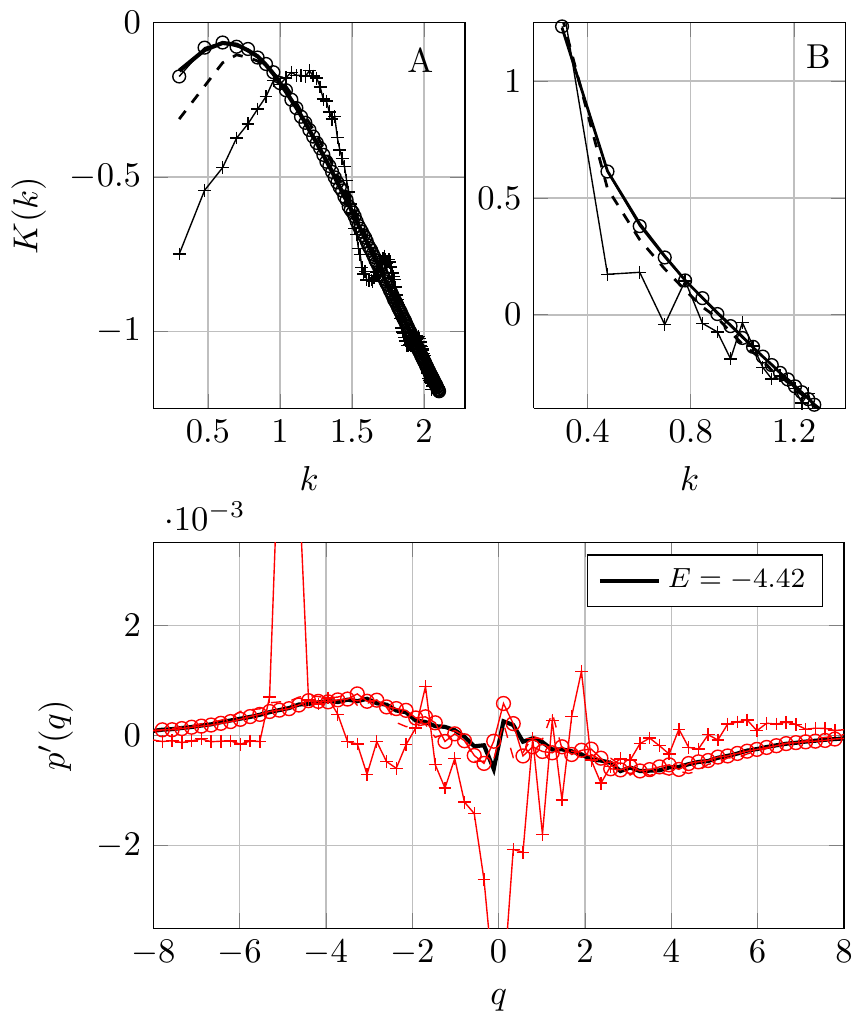}
\end{center}
\vspace{-0.25in}
\caption{Top panels: Evolution of the kinetic energy spectra, $K(k)$ for
(A) low-energy, positive temperature ($E =-4.42$) and (B) high-energy, negative
temperature ($E = 1.66$). The initial spectra are shown by $+$, the long time
spectra by $\circ$ and the microcanonical estimate (from Fig. 2) is bold. Dashed lines
indicate results for short times. Bottom panel: Evolution of residual probability $p'_{int}$
for case A. Symbols are the same as above.}
\label{fig4a}
\end{figure}

As a 
yet
stronger test of ergodicity, we consider the evolution of states with
\textit{atypical} initial spectra for a given energy.
First,
an ensemble of 111 states was generated in the  
strongly positive temperature ($E \sim -4.42$) system by
randomly placing vortex dipoles (opposite signed pairs separated by $\bar{d}/\sqrt{2}$) 
instead of single vortices.
For such dipole states, the
kinetic energy spectrum $\langle K \rangle(k)$ (averaged over the 111 states),
shown by the $+$ symbols in Fig.~\ref{fig4a}A, 
differs significantly (beyond several microcanonical standard deviations) from
the microcanonical mean (thick solid line).  
However, 
upon evolution the dipole initial states rapidly relax
towards the microcanonical mean. The dipole spectrum time averaged over $2 \le t \le 4$
is shown by the dashed line, and the late time-averaged spectrum
($392 \le t \le 400$, open circles) 
%
is statistically indistinguishable from the microcanonical estimate.
In addition, the standard deviation in the spectrum also 
converges 
to that of the  microcanonical ensemble (not shown).

\begin{figure}[!htb]
\begin{center}
{\includegraphics[width=0.25\textwidth]{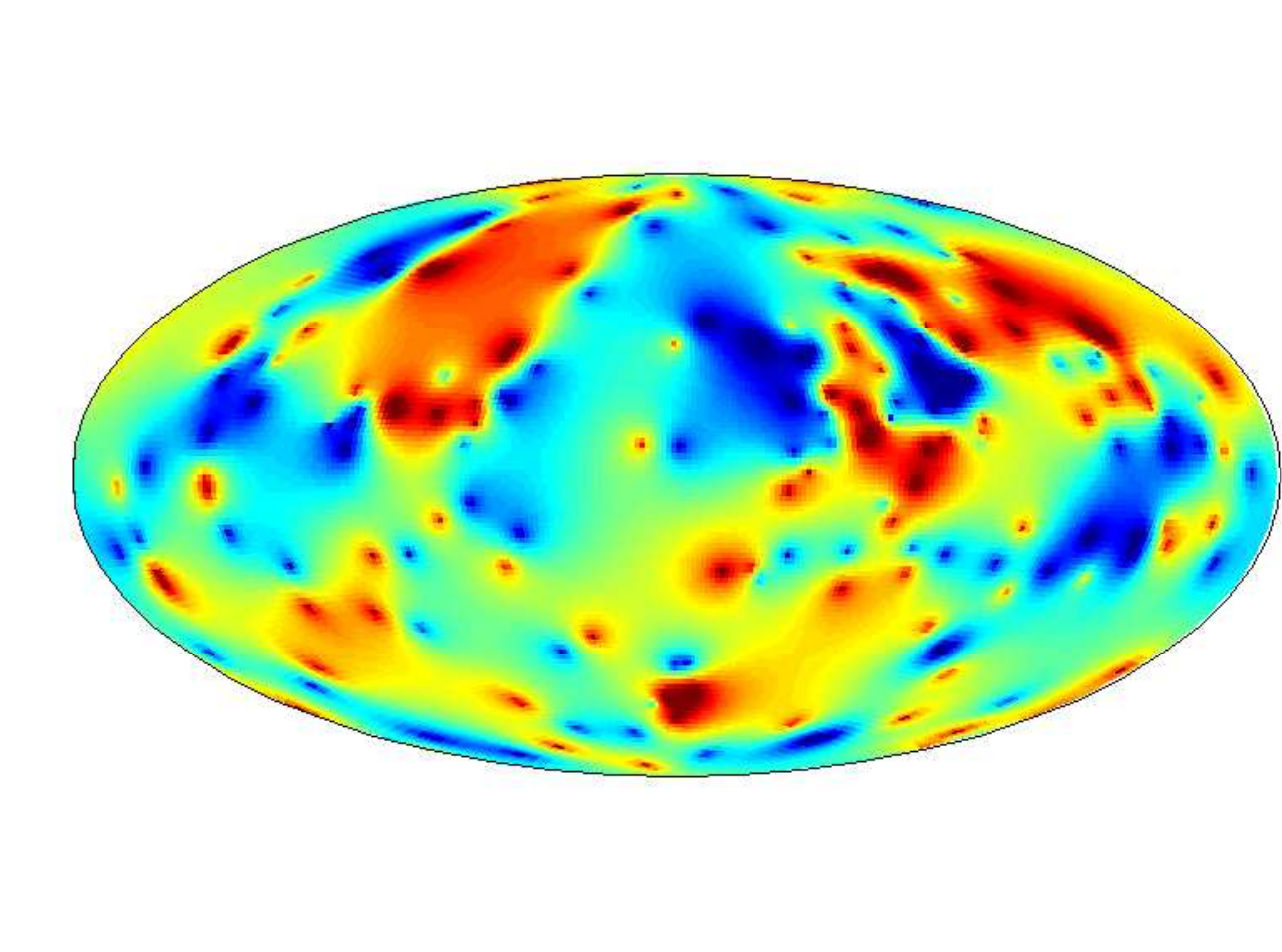}}
\put(-118,80){(a1)}
{\includegraphics[width=0.25\textwidth]{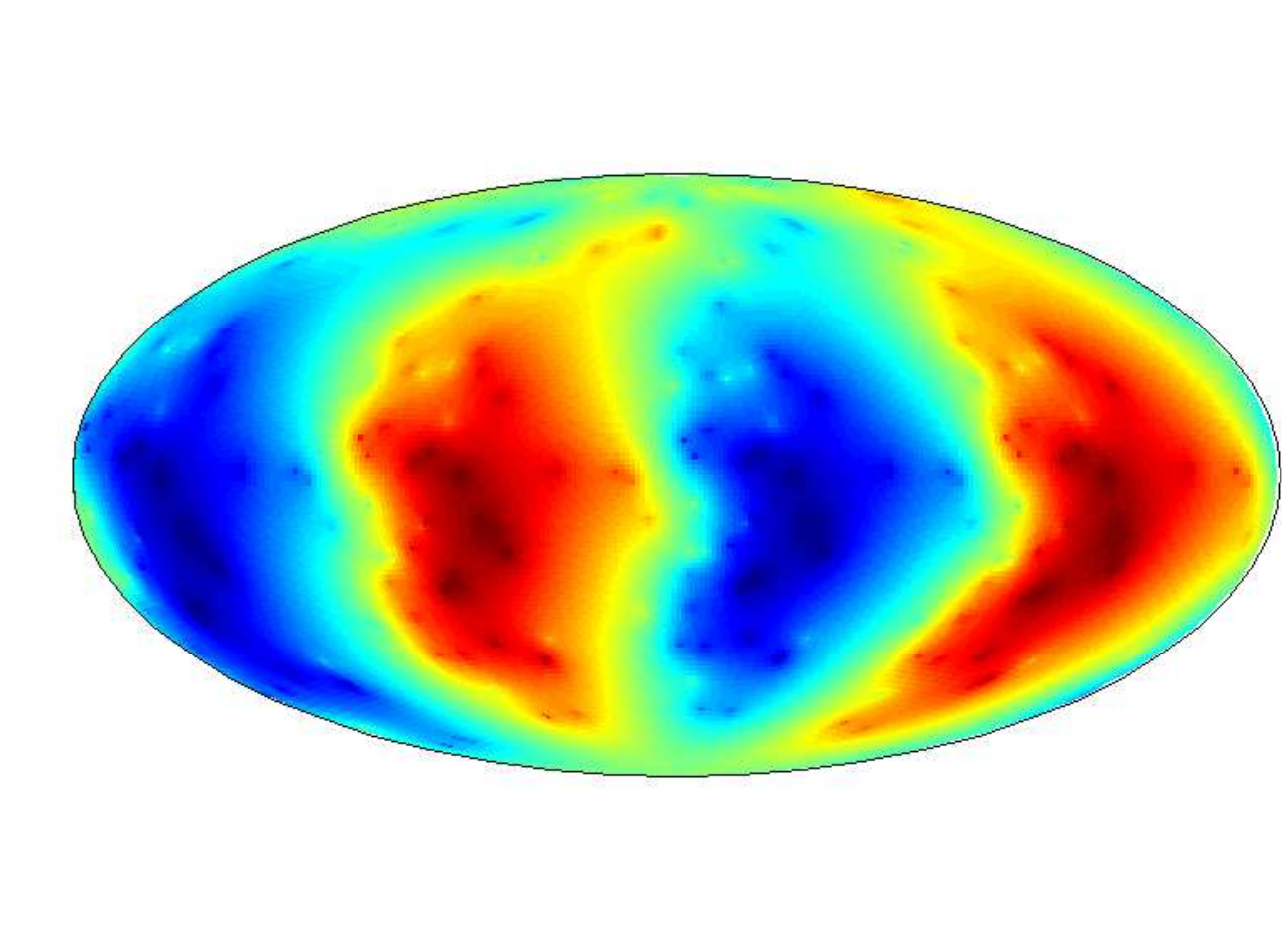}}
\put(-115,80){(b1)}
\vspace{-0.4cm}

{\includegraphics[width=0.25\textwidth]{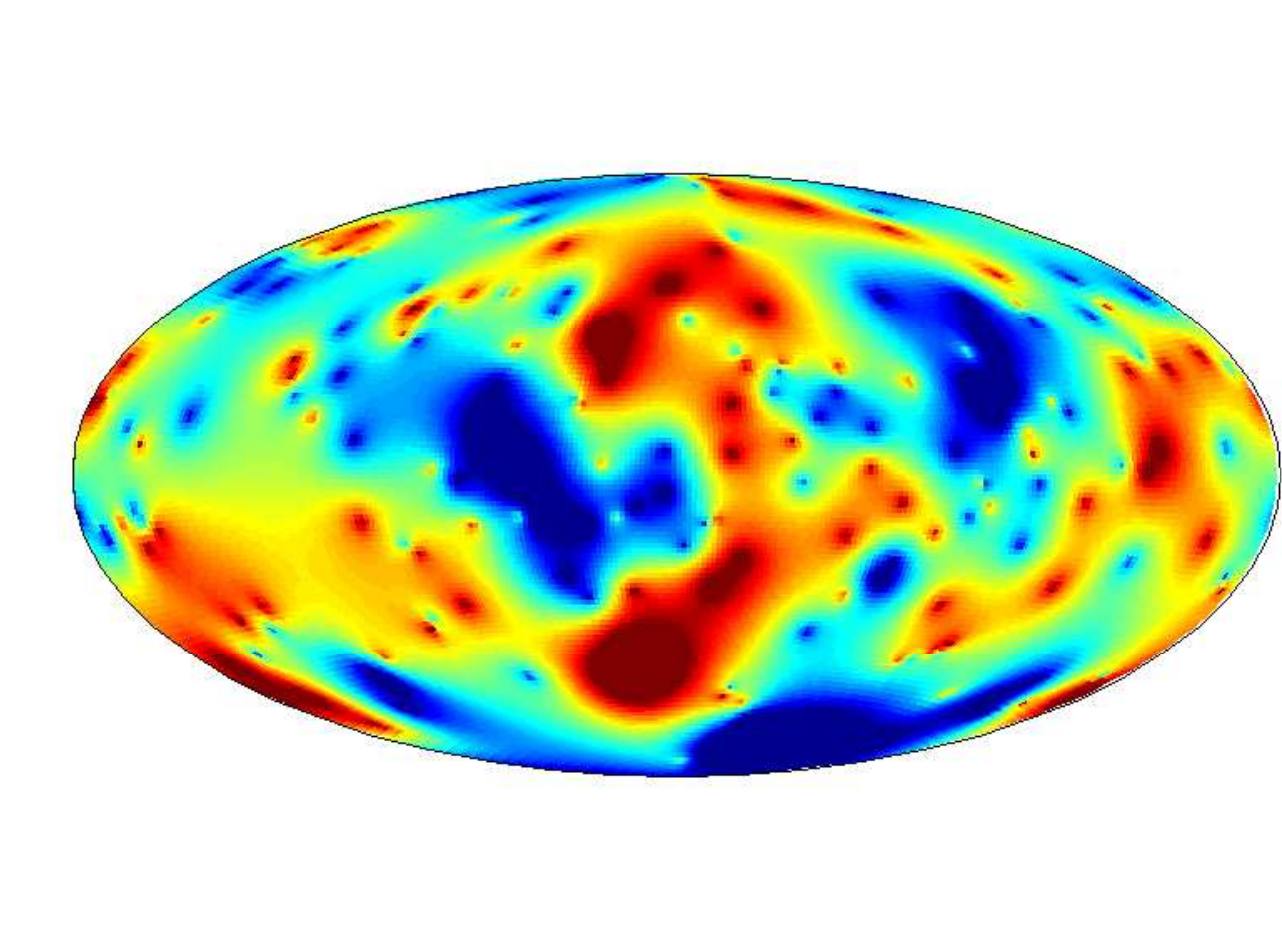}}
\put(-118,80){(a2)}
{\includegraphics[width=0.25\textwidth]{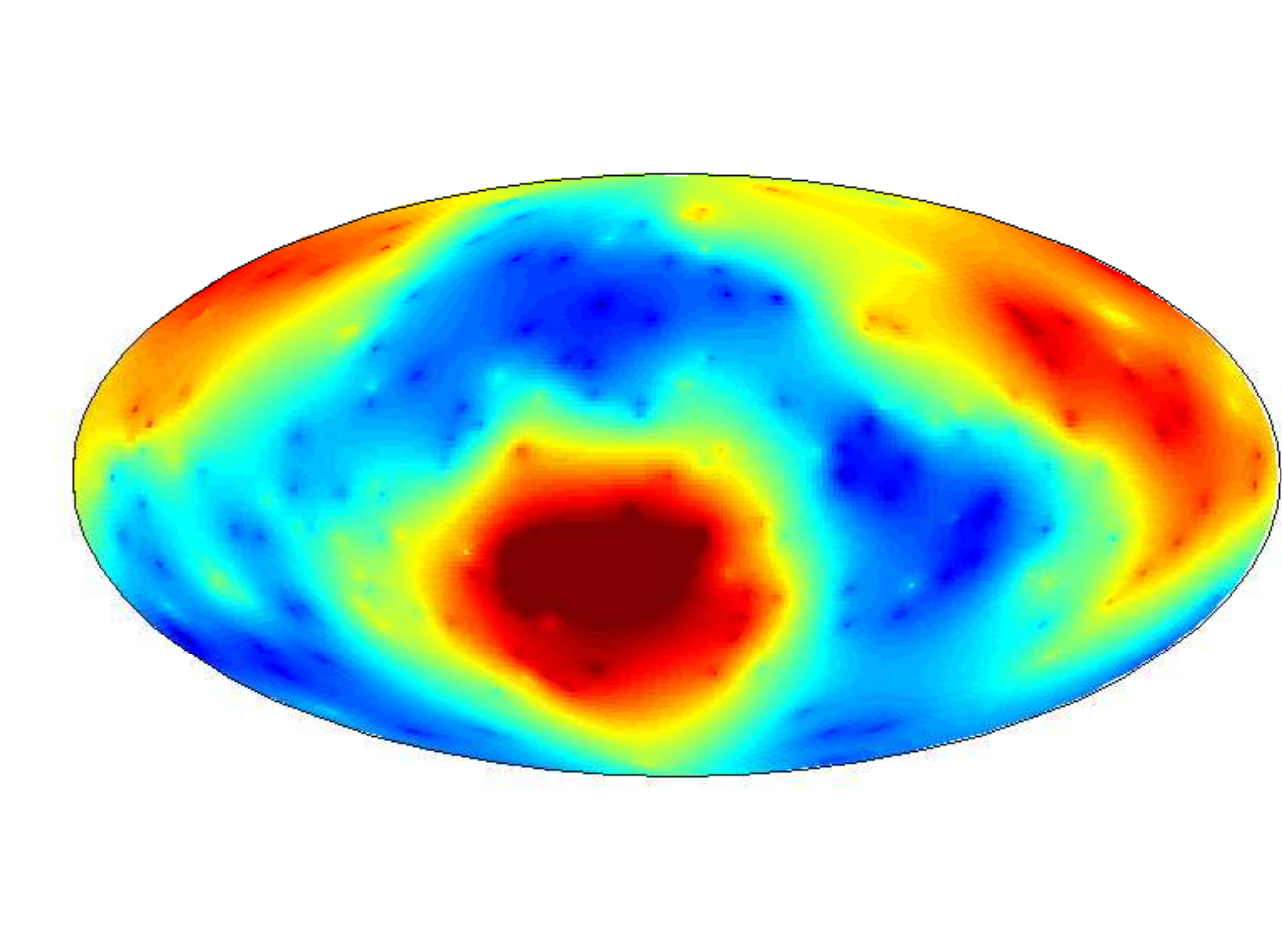}}
\put(-118,80){(b2)}
\vspace{-0.4cm}

{\includegraphics[width=0.25\textwidth]{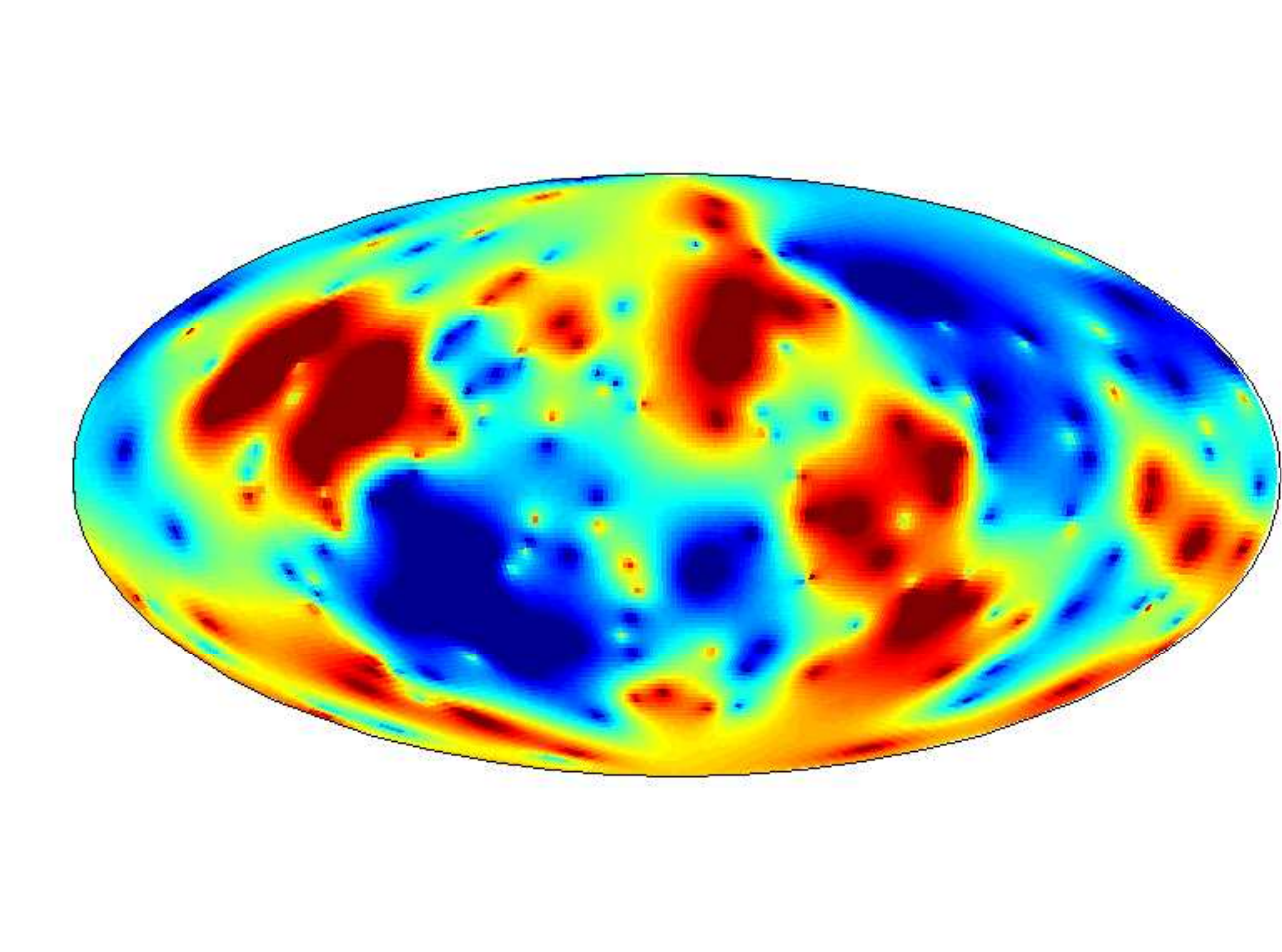}}
\put(-118,80){(a3)}
{\includegraphics[width=0.25\textwidth]{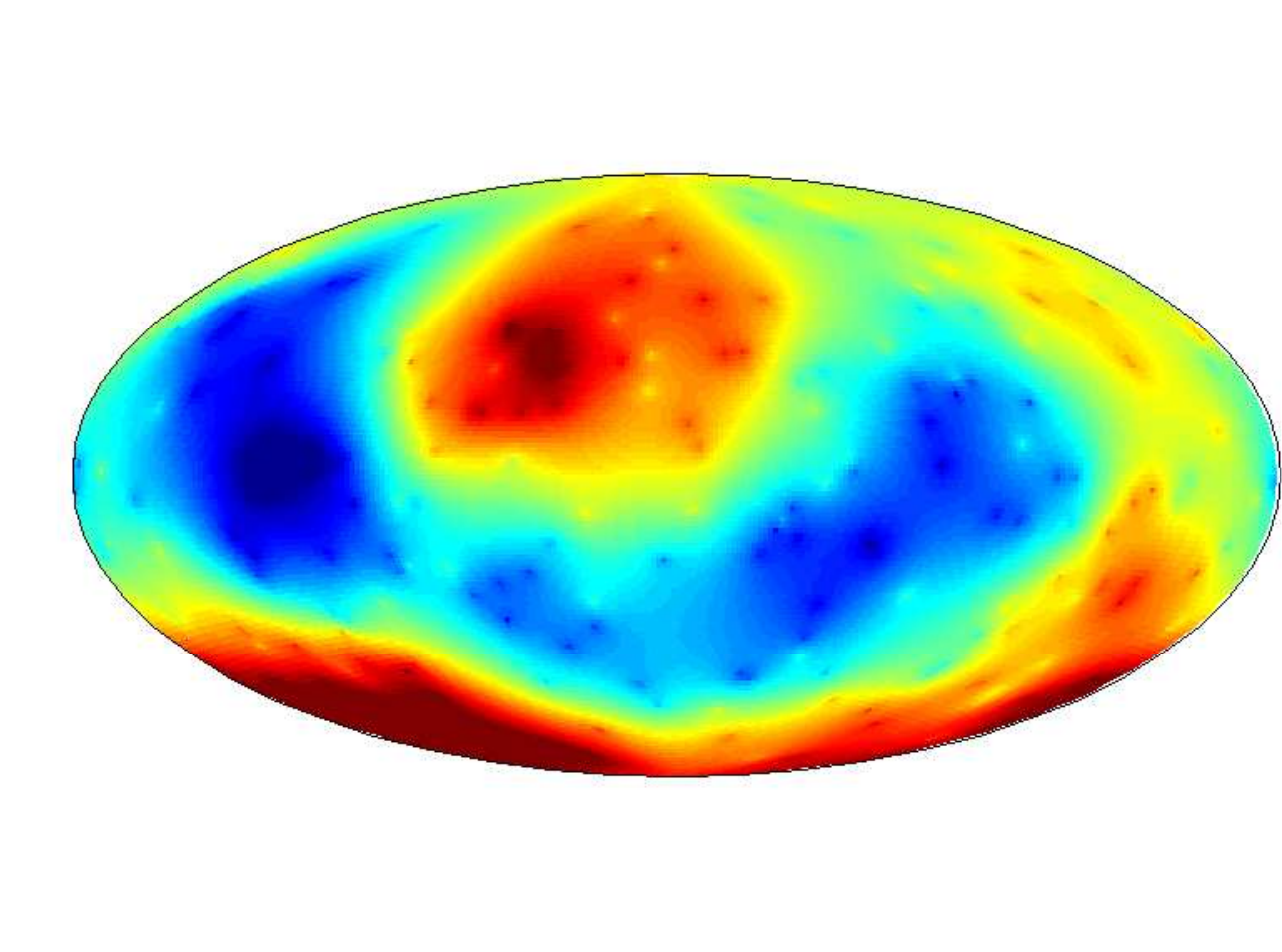}}
\put(-118,80){(b3)}
\end{center}
\vspace{-0.25in}
\caption{ (Color online) Left panels: Evolution of the dipole streamfunction
  at $E = -4.42$.  (a1)
  Initial dipole streamfunction. (a2) dipole streamfunction at 
  $t = 400$. (a3) Initial streamfunction for a representative ensemble member
  at the same energy. Right panels: Same as left but for forward
  cascade case, $E = 1.66$.  The projection shows the entire sphere
  and the color scale is constant for each energy.}
\label{fig5}
\end{figure}

Vortex interactions immediately destroy the initial
equal vortex-pair separation, and the distribution of pair separations
continues to spread until the state resembles a randomly chosen
collection of vortices for this energy.  
As shown in the lower panel of  Fig.~\ref{fig4a},  the initial residual probability $p'_{\rm int}(q)$  spikes
at the $q$ value of the dipole separation, but then relaxes to the microcanonical estimate (open
circles show the late time average). 
This relaxation can be seen directly in the streamfunction of any dipole initial condition.
The left panels of fig.~\ref{fig5} shows the evolution of $\psi(\theta,\phi)$ from an initial dipole state (a1)
to  $t=400$ (a2) along with the streamfunction of a randomly chosen member of the microcanonical 
ensemble (a3).
For this positive temperature state, there is an inverse cascade of kinetic energy to large scales. 

Similar results have been found starting from atypical states in 
the highest energy range, $E \sim 1.66$ where the temperature is negative.  
By randomly placing vortices with
an increased probability to project on the $k=2$ spherical harmonic, 
a surplus of kinetic energy is created at the 
largest permissible scale for  $J=0$.  As seen in Fig.~\ref{fig4a}B,
the initial $\langle K \rangle(k)$ ($+$ symbols) again rapidly relaxes
back to the microcanonical estimate (bold line) with the dashed line
showing the spectrum at times $2 \le t \le 4$ and the open circles the late time spectrum.  
Corresponding behavior in real space for an individual initial condition
is shown in the right column of Fig.~\ref{fig5}, with an initial atypical state 
in (b1) 
(the
pattern closely matches a spherical harmonic), the same state at 
$t=400$ (b2), and a randomly selected member of the 
microcanonical ensemble (b3).
The images in (b2) and (b3) exhibit more smaller-scale features than the image
in (b1) and, as shown in the spectral evolution, there is a forward cascade of kinetic
energy despite the negative system temperature.

The relaxation of atypical states to the microcanonical average occurs
on a {\it short} timescale, comparable to the rotation period $\tau$
of two like-signed vortices separated by a distance $\bar{d}$ (also
the time taken for a dipole pair to propagate a distance $\bar{d}$).
That is, any special order in the initial conditions is rapidly
destroyed by the ensuing dynamical evolution.  This is a strong
indication of ergodic dynamics in this geometry, and appears to
contrast with the results of \cite{wm91} in doubly-periodic geometry.
On the other hand, \cite{wm91} considered just 6 vortices.  It might
be that so few vortices have insufficient freedom to fully randomize
and resemble typical, microcanonical states.

To test this, we repeated the analysis above for $n=8$ vortices in
spherical geometry.  After accounting for additional constraints
arising from conservation of angular impulse ${\mathbf I}$,
the two systems have a similar number of degrees of freedom.
Here, we focus on the evolution of atypical low energy states,
with $E \approx -4.42$.  These states were generated by placing 
4 pairs of dipoles at random, with the halves of each pair 
separated by $d = 2e^{E/2} \approx 0.22$.  At this distance,
the individual energies of the dipoles sum to $E$.  The 
additional energy contributed by inter-pair interactions is
${\mathcal{O}}(d^2/\bar{d}^2) \ll 1$ and is easily canceled
by appropriate placement of the pairs.  

\begin{figure}[!htb]
\begin{center}
    \includegraphics[width=0.49\textwidth]{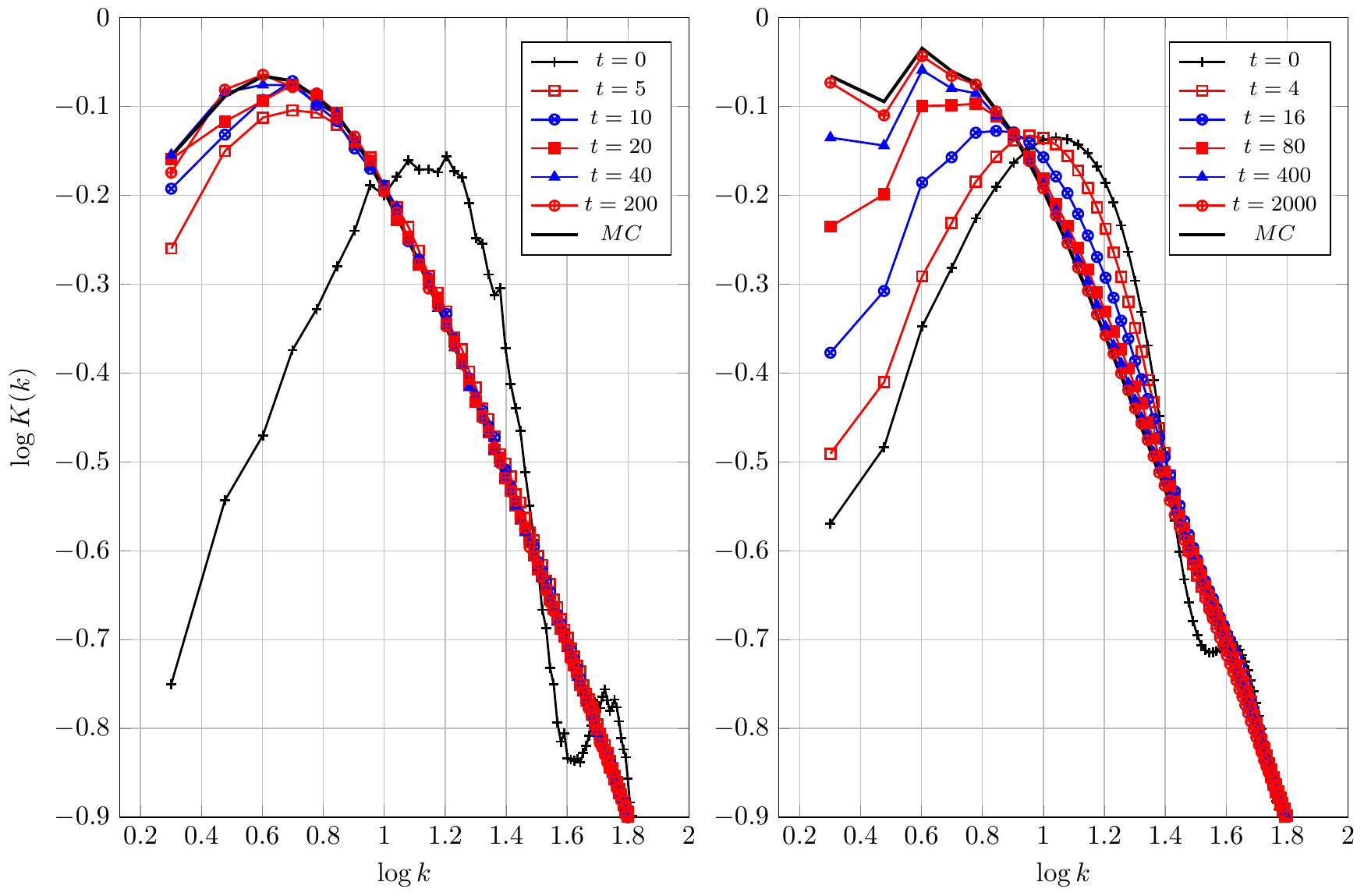} \\
    \includegraphics[width=0.49\textwidth]{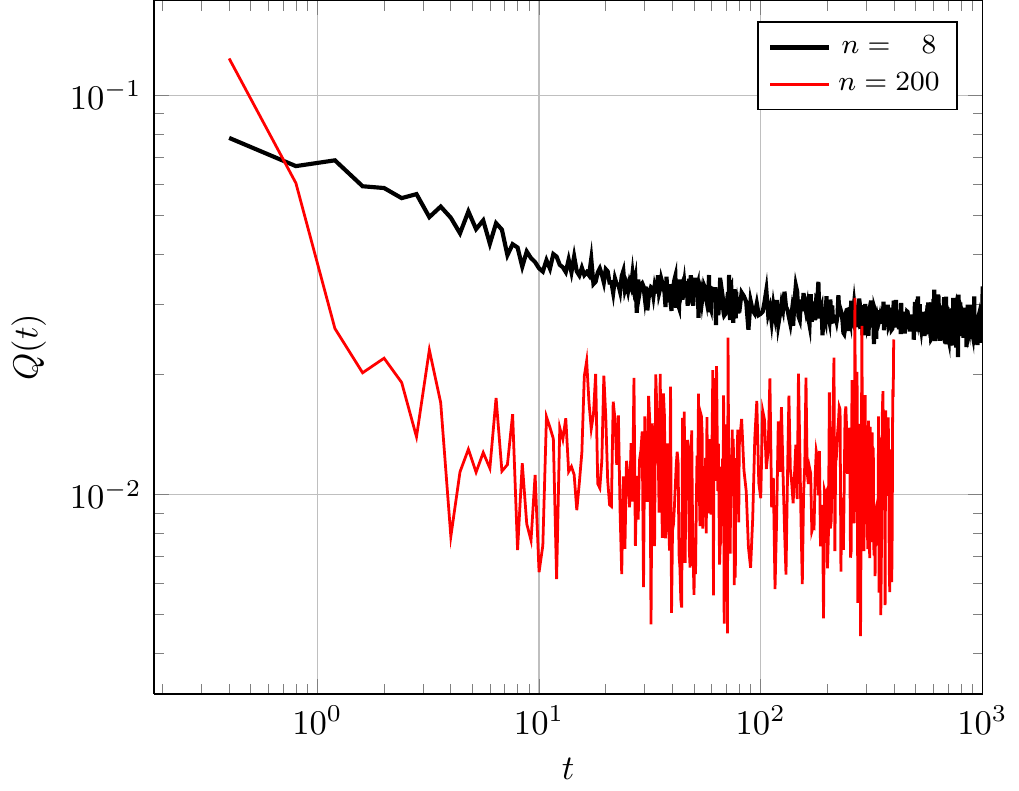} \\
\end{center}
\vspace{-0.25in}
\caption{ 
Evolution of the kinetic energy spectrum for dipole initial conditions at $E = -4.42$ for $n=200$ (upper left panel)
and $n = 8$ (upper right panel). Lower panel shows the temporal evolution of the $Q$ measure for the two cases.}
\label{fig6}
\end{figure}

The upper panels of Fig.~\ref{fig6} contrast the ensemble-averaged
spectral evolution of atypical states for $n=200$ (seen before) on the
left with that for $n=8$ on the right.  For $n=8$, the ensemble
consists of 1000 states (an increase on the 111 states used for
$n=200$ to reduce the variance in $K(k)$).  Both systems
clearly show spectral relaxation to the microcanonical mean. The 
relaxation rate, however, is much slower in the dilute, $n=8$ case.

Another measure of relaxation is obtained by analyzing the
probability distribution $p_{\rm int}(q)$ of the normalized pairwise
interaction energies in \eqref{eq:IJInteraction}.  Denote
$\bar{p}_{\rm int}(q)$ as the microcanonical ensemble mean, and
$\bar{\delta}_{\rm int}$ as the integrated standard deviation of
individual members of the ensemble from $\bar{p}_{\rm int}(q)$.  We
measure the relaxation of the dynamical ensemble to the
microcanonical one by 
$Q(t) \equiv {\delta}_{\rm int}(t)/\bar{\delta}_{\rm int}$ 
where ${\delta}_{\rm int}$ is the integrated standard deviation of
individual members of the dynamical ensemble from 
$\bar{p}_{\rm int}(q)$.  This measure is shown in the bottom panel
of Fig.~\ref{fig6} (note logarithmic scales).  Consistent with the
spectral evolution, $Q(t)$ for $n=200$ decreases rapidly to a 
low, fluctuating level (the logarithmic scaling makes this 
fluctuation appear much larger than it actually is).  By contrast,
the decay of $Q(t)$ is much slower for $n=8$, though nonetheless 
it approaches a roughly constant level at late times.  The higher
equilibrated level for $n=8$ is predominantly due to differences
in the statistical sample size.  The calculation of $p_{\rm int}(q)$
involves $n(n-1)/2$ separate vortex interactions.
This is substantially larger for $n=200$ than for $n=8$, and while 9 times
more cases were considered for $n=8$, the sample size is still 
approximately 80 times larger for $n=200$ than for $n=8$.

The results show that atypical dynamical states inevitably relax to
the equilibrium microcanonical distribution, independent of the system
size (at least for $n \geq 8$).  Equilibration is observed in both the
kinetic energy spectra $K(k)$ and in the complementary measure $Q$
using $p_{\rm int}(q)$.

Although relaxation is observed both for $n=8$ and $n=200$, there is a
striking difference in the rate of relaxation in the two cases.
For $n=200$, we have found that the relaxation occurs on the
characteristic timescale $\tau$, whereas for $n=8$ it is considerably
slower. 
Given that the circulations are scaled by $\sqrt{n}$, the
characteristic timescale is independent of system size. Therefore, the
observed difference in relaxation timescales cannot be explained
simply by differences between dipole collision rates in the two
systems.

\begin{figure}[!htb]
\begin{center}
    \includegraphics[width=0.49\textwidth]{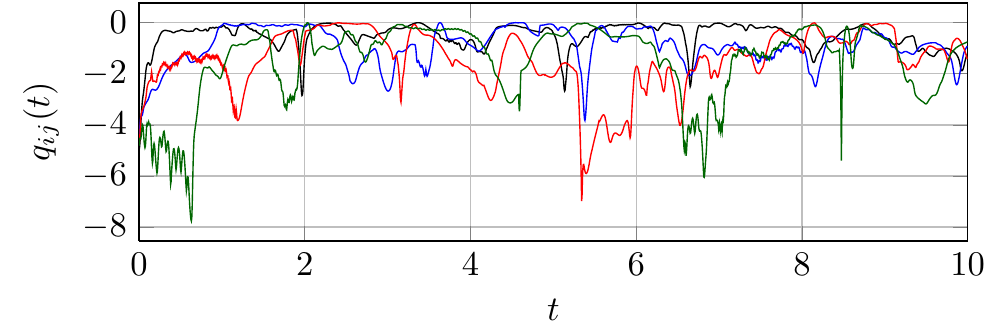} \\
    \includegraphics[width=0.49\textwidth]{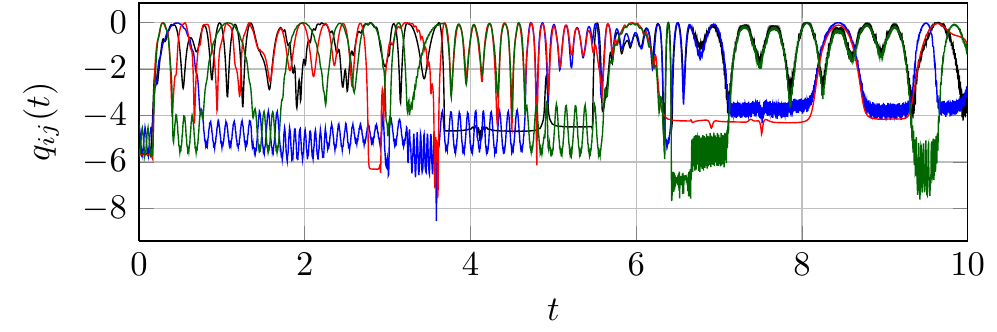} 
\end{center}
\vspace{-0.25in}
\caption{ 
Evolution of the pairwise energy for (top panel) the 4 initial dipoles for $n=8$ 
and (lower panel) 4 initial dipoles in the $n = 200$ case.}
\label{fig7}
\end{figure}

Movies of the vortex motion in the dilute dipole case indicate that a
majority of dipole interactions involve only simple particle exchange,
producing no discernible change in the separation of the vortices in
each pair.  Such interactions preserve the structure of the initial
conditions and rapid statistical evolution requires higher-order
collisions involving interactions between three or more
dipole pairs.
Evidence that such interactions occur far less frequently in
the $n=8$ case is provided in Fig.~\ref{fig7}.  Here we show the 
early time evolution of
the pairwise energy, $q_{ij}$, of 4 initial dipoles --- the complete set
when $n=8$, and 4 randomly selected from the 100 available when $n=200$.
Particle exchange collisions are clearly evident in the dilute case
($n=8$, upper panel) where individual pair energies spike to zero before
consistently returning to the negative energy level associated with
their initial separations.  Two dipoles pairs repeat this process more
than 5 times in the first 10 time units.  The time distribution of
pairwise energies is bimodal, highly concentrated at $-|q_0|$ and
0. By contrast, initial pairs in the $n=200$ case are rapidly
scattered and information about the interaction energy of the initial
configuration is quickly lost.  This is true not only for the 4
selected dipole pairs shown here, but for all pairs in the $n=200$ case.



\medskip

\section{Conclusions}

Due to the universal $k^{-1}$ behavior of point-vortex kinetic energy spectra at small scales,
increasing the system energy preferentially increases the kinetic energy content at the
largest allowable scales. While this is entirely consistent with Onsager's conjecture concerning 
the increased likelihood of observing large-scale structure at sufficiently high energies, notably 
it is also independent of the thermodynamic temperature of the system.  
In addition, the results indicate that point-vortex dynamics, at least on the isotropic
sphere, are ergodic and therefore statistical measures derived from the
dynamics of almost all initial states simply relax to those given by the
microcanonical ensemble. For the kinetic energy spectra (equivalently
$p_{int}(q)$ distributions) examined here, the relaxation takes place
on timescales comparable to an eddy turnover time, independent of the system temperature. 
As such, for the simplest bounded domain, there is no 
direct relationship between the sign of
 the statistical temperature and the
direction of any dynamic cascade process in the velocity field induced by a finite number of point vortices.

\smallskip
ACP supported under DOD (MURI) grant
N000141110087 ONR.  The computations were supported by the CUNY HPCC 
under NSF Grants CNS-0855217 and CNS-0958379. The authors thank C. Lancellotti
for fruitful discussions.


\begin{thebibliography}{25}%
\makeatletter
\providecommand \@ifxundefined [1]{%
 \@ifx{#1\undefined}
}%
\providecommand \@ifnum [1]{%
 \ifnum #1\expandafter \@firstoftwo
 \else \expandafter \@secondoftwo
 \fi
}%
\providecommand \@ifx [1]{%
 \ifx #1\expandafter \@firstoftwo
 \else \expandafter \@secondoftwo
 \fi
}%
\providecommand \natexlab [1]{#1}%
\providecommand \enquote  [1]{``#1''}%
\providecommand \bibnamefont  [1]{#1}%
\providecommand \bibfnamefont [1]{#1}%
\providecommand \citenamefont [1]{#1}%
\providecommand \href@noop [0]{\@secondoftwo}%
\providecommand \href [0]{\begingroup \@sanitize@url \@href}%
\providecommand \@href[1]{\@@startlink{#1}\@@href}%
\providecommand \@@href[1]{\endgroup#1\@@endlink}%
\providecommand \@sanitize@url [0]{\catcode `\\12\catcode `\$12\catcode
  `\&12\catcode `\#12\catcode `\^12\catcode `\_12\catcode `\%12\relax}%
\providecommand \@@startlink[1]{}%
\providecommand \@@endlink[0]{}%
\providecommand \url  [0]{\begingroup\@sanitize@url \@url }%
\providecommand \@url [1]{\endgroup\@href {#1}{\urlprefix }}%
\providecommand \urlprefix  [0]{URL }%
\providecommand \Eprint [0]{\href }%
\providecommand \doibase [0]{http://dx.doi.org/}%
\providecommand \selectlanguage [0]{\@gobble}%
\providecommand \bibinfo  [0]{\@secondoftwo}%
\providecommand \bibfield  [0]{\@secondoftwo}%
\providecommand \translation [1]{[#1]}%
\providecommand \BibitemOpen [0]{}%
\providecommand \bibitemStop [0]{}%
\providecommand \bibitemNoStop [0]{.\EOS\space}%
\providecommand \EOS [0]{\spacefactor3000\relax}%
\providecommand \BibitemShut  [1]{\csname bibitem#1\endcsname}%
\let\auto@bib@innerbib\@empty
\bibitem [{\citenamefont {Kirchhoff}(1876)}]{Kirchhoff:1876}%
  \BibitemOpen
  \bibfield  {author} {\bibinfo {author} {\bibfnamefont {G.}~\bibnamefont
  {Kirchhoff}},\ }\href@noop {} {\emph {\bibinfo {title} {Vorlesungen \"uber
  mathematische Physik}}}\ (\bibinfo  {publisher} {Teubner, Leipzig},\ \bibinfo
  {year} {1876})\BibitemShut {NoStop}%
\bibitem [{\citenamefont {Suryanarayanan}\ \emph {et~al.}(2014)\citenamefont
  {Suryanarayanan}, \citenamefont {Narasimha},\ and\ \citenamefont
  {Dass}}]{PhysRevE.89.013009}%
  \BibitemOpen
  \bibfield  {author} {\bibinfo {author} {\bibfnamefont {S.}~\bibnamefont
  {Suryanarayanan}}, \bibinfo {author} {\bibfnamefont {R.}~\bibnamefont
  {Narasimha}}, \ and\ \bibinfo {author} {\bibfnamefont {N.~D.~H.}\
  \bibnamefont {Dass}},\ }\href {\doibase 10.1103/PhysRevE.89.013009}
  {\bibfield  {journal} {\bibinfo  {journal} {Phys. Rev. E}\ }\textbf {\bibinfo
  {volume} {89}},\ \bibinfo {pages} {013009} (\bibinfo {year}
  {2014})}\BibitemShut {NoStop}%
\bibitem [{\citenamefont {Wang}\ \emph {et~al.}(2007)\citenamefont {Wang},
  \citenamefont {Sergeev}, \citenamefont {Barenghi},\ and\ \citenamefont
  {Harrison}}]{Wang2007}%
  \BibitemOpen
  \bibfield  {author} {\bibinfo {author} {\bibfnamefont {S.}~\bibnamefont
  {Wang}}, \bibinfo {author} {\bibfnamefont {Y.}~\bibnamefont {Sergeev}},
  \bibinfo {author} {\bibfnamefont {C.}~\bibnamefont {Barenghi}}, \ and\
  \bibinfo {author} {\bibfnamefont {M.}~\bibnamefont {Harrison}},\ }\href
  {\doibase 10.1007/s10909-007-9499-2} {\bibfield  {journal} {\bibinfo
  {journal} {J. Low Temp. Phys.}\ }\textbf {\bibinfo {volume} {149}},\ \bibinfo
  {pages} {65} (\bibinfo {year} {2007})}\BibitemShut {NoStop}%
\bibitem [{\citenamefont {Simula}\ \emph {et~al.}(2014)\citenamefont {Simula},
  \citenamefont {Davis},\ and\ \citenamefont {Helmerson}}]{simula}%
  \BibitemOpen
  \bibfield  {author} {\bibinfo {author} {\bibfnamefont {T.}~\bibnamefont
  {Simula}}, \bibinfo {author} {\bibfnamefont {M.~J.}\ \bibnamefont {Davis}}, \
  and\ \bibinfo {author} {\bibfnamefont {K.}~\bibnamefont {Helmerson}},\
  }\href@noop {} {\bibfield  {journal} {\bibinfo  {journal} {Phys. Rev. Lett.}\
  }\textbf {\bibinfo {volume} {113}},\ \bibinfo {pages} {165302} (\bibinfo
  {year} {2014})}\BibitemShut {NoStop}%
\bibitem [{\citenamefont {Kiessling}\ and\ \citenamefont
  {Lebowitz}(1997)}]{Kiessling97}%
  \BibitemOpen
  \bibfield  {author} {\bibinfo {author} {\bibfnamefont {M.-H.}\ \bibnamefont
  {Kiessling}}\ and\ \bibinfo {author} {\bibfnamefont {J.}~\bibnamefont
  {Lebowitz}},\ }\href@noop {} {\bibfield  {journal} {\bibinfo  {journal}
  {Lett. Math. Phys.}\ }\textbf {\bibinfo {volume} {42}},\ \bibinfo {pages}
  {43} (\bibinfo {year} {1997})}\BibitemShut {NoStop}%
\bibitem [{\citenamefont {Chavanis}(2012)}]{Chavanis12}%
  \BibitemOpen
  \bibfield  {author} {\bibinfo {author} {\bibfnamefont {P.-H.}\ \bibnamefont
  {Chavanis}},\ }\href {\doibase 10.1016/j.physa.2012.02.014} {\bibfield
  {journal} {\bibinfo  {journal} {Phys. A}\ }\textbf {\bibinfo {volume}
  {391}},\ \bibinfo {pages} {3657} (\bibinfo {year} {2012})}\BibitemShut
  {NoStop}%
\bibitem [{\citenamefont {Onsager}(1949)}]{Onsager}%
  \BibitemOpen
  \bibfield  {author} {\bibinfo {author} {\bibfnamefont {L.}~\bibnamefont
  {Onsager}},\ }\href@noop {} {\bibfield  {journal} {\bibinfo  {journal} {Nuovo
  Cimento}\ }\textbf {\bibinfo {volume} {6}},\ \bibinfo {pages} {279} (\bibinfo
  {year} {1949})}\BibitemShut {NoStop}%
\bibitem [{\citenamefont {Eyink}\ and\ \citenamefont
  {Sreenivasan}(2006)}]{EyinkSree:2006}%
  \BibitemOpen
  \bibfield  {author} {\bibinfo {author} {\bibfnamefont {G.}~\bibnamefont
  {Eyink}}\ and\ \bibinfo {author} {\bibfnamefont {K.}~\bibnamefont
  {Sreenivasan}},\ }\href@noop {} {\bibfield  {journal} {\bibinfo  {journal}
  {Rev. Modern Phys.}\ }\textbf {\bibinfo {volume} {70}},\ \bibinfo {pages}
  {87} (\bibinfo {year} {2006})}\BibitemShut {NoStop}%
\bibitem [{\citenamefont {Montgomery}\ and\ \citenamefont
  {Joyce}(1974)}]{MontgomeryJoyce}%
  \BibitemOpen
  \bibfield  {author} {\bibinfo {author} {\bibfnamefont {D.}~\bibnamefont
  {Montgomery}}\ and\ \bibinfo {author} {\bibfnamefont {G.}~\bibnamefont
  {Joyce}},\ }\href@noop {} {\bibfield  {journal} {\bibinfo  {journal} {Phys.
  Fluids}\ }\textbf {\bibinfo {volume} {17}},\ \bibinfo {pages} {1139}
  (\bibinfo {year} {1974})}\BibitemShut {NoStop}%
\bibitem [{\citenamefont {Lundgren}\ and\ \citenamefont
  {Pointin}(1977)}]{LundPoint}%
  \BibitemOpen
  \bibfield  {author} {\bibinfo {author} {\bibfnamefont {T.}~\bibnamefont
  {Lundgren}}\ and\ \bibinfo {author} {\bibfnamefont {Y.}~\bibnamefont
  {Pointin}},\ }\href@noop {} {\bibfield  {journal} {\bibinfo  {journal} {J.
  Stat. Phys.}\ }\textbf {\bibinfo {volume} {17}},\ \bibinfo {pages} {323}
  (\bibinfo {year} {1977})}\BibitemShut {NoStop}%
\bibitem [{\citenamefont {Kraichnan}\ and\ \citenamefont
  {Montgomery}(1980)}]{KraichnanMontgomery}%
  \BibitemOpen
  \bibfield  {author} {\bibinfo {author} {\bibfnamefont {R.}~\bibnamefont
  {Kraichnan}}\ and\ \bibinfo {author} {\bibfnamefont {D.}~\bibnamefont
  {Montgomery}},\ }\href@noop {} {\bibfield  {journal} {\bibinfo  {journal}
  {Rep. Prog. Phys}\ }\textbf {\bibinfo {volume} {43}},\ \bibinfo {pages} {547}
  (\bibinfo {year} {1980})}\BibitemShut {NoStop}%
\bibitem [{\citenamefont {Miller}(1990)}]{Miller}%
  \BibitemOpen
  \bibfield  {author} {\bibinfo {author} {\bibfnamefont {J.}~\bibnamefont
  {Miller}},\ }\href@noop {} {\bibfield  {journal} {\bibinfo  {journal} {Phys.
  Rev. Lett.}\ }\textbf {\bibinfo {volume} {65}},\ \bibinfo {pages} {2137}
  (\bibinfo {year} {1990})}\BibitemShut {NoStop}%
\bibitem [{\citenamefont {Robert}\ and\ \citenamefont
  {Sommeria}(1992)}]{Robert}%
  \BibitemOpen
  \bibfield  {author} {\bibinfo {author} {\bibfnamefont {R.}~\bibnamefont
  {Robert}}\ and\ \bibinfo {author} {\bibfnamefont {J.}~\bibnamefont
  {Sommeria}},\ }\href@noop {} {\bibfield  {journal} {\bibinfo  {journal}
  {Phys. Rev. Lett.}\ }\textbf {\bibinfo {volume} {69}},\ \bibinfo {pages}
  {2776} (\bibinfo {year} {1992})}\BibitemShut {NoStop}%
\bibitem [{\citenamefont {B\"uhler}(2002)}]{Buhler}%
  \BibitemOpen
  \bibfield  {author} {\bibinfo {author} {\bibfnamefont {O.}~\bibnamefont
  {B\"uhler}},\ }\href@noop {} {\bibfield  {journal} {\bibinfo  {journal}
  {Phys. Fluids}\ }\textbf {\bibinfo {volume} {14}},\ \bibinfo {pages} {2139}
  (\bibinfo {year} {2002})}\BibitemShut {NoStop}%
\bibitem [{\citenamefont {Yatsuyanagi}(2005)}]{yatsu}%
  \BibitemOpen
  \bibfield  {author} {\bibinfo {author} {\bibfnamefont {Y.}~\bibnamefont
  {Yatsuyanagi}},\ }\href@noop {} {\bibfield  {journal} {\bibinfo  {journal}
  {Phys. Rev. Lett.}\ }\textbf {\bibinfo {volume} {94}},\ \bibinfo {pages}
  {0544502} (\bibinfo {year} {2005})}\BibitemShut {NoStop}%
\bibitem [{\citenamefont {McWilliams}(1983)}]{Mac83}%
  \BibitemOpen
  \bibfield  {author} {\bibinfo {author} {\bibfnamefont {J.}~\bibnamefont
  {McWilliams}},\ }\href@noop {} {\bibfield  {journal} {\bibinfo  {journal} {J.
  Fluid Mech.}\ }\textbf {\bibinfo {volume} {146}},\ \bibinfo {pages} {21}
  (\bibinfo {year} {1983})}\BibitemShut {NoStop}%
\bibitem [{\citenamefont {Dritschel}\ \emph {et~al.}(2008)\citenamefont
  {Dritschel}, \citenamefont {Scott}, \citenamefont {R.K.}, \citenamefont
  {MacAskill}, \citenamefont {Gottwald},\ and\ \citenamefont {C.V.Tran}}]{d08}%
  \BibitemOpen
  \bibfield  {author} {\bibinfo {author} {\bibfnamefont {D.}~\bibnamefont
  {Dritschel}}, \bibinfo {author} {\bibfnamefont {R.}~\bibnamefont {Scott}},
  \bibinfo {author} {\bibnamefont {R.K.}}, \bibinfo {author} {\bibfnamefont
  {C.}~\bibnamefont {MacAskill}}, \bibinfo {author} {\bibfnamefont
  {G.}~\bibnamefont {Gottwald}}, \ and\ \bibinfo {author} {\bibnamefont
  {C.V.Tran}},\ }\href@noop {} {\bibfield  {journal} {\bibinfo  {journal}
  {Phys. Rev. Lett.}\ }\textbf {\bibinfo {volume} {101}},\ \bibinfo {pages}
  {094501} (\bibinfo {year} {2008})}\BibitemShut {NoStop}%
\bibitem [{\citenamefont {Eyink}\ and\ \citenamefont
  {Spohn}(1993)}]{EyinkSpohn:1993}%
  \BibitemOpen
  \bibfield  {author} {\bibinfo {author} {\bibfnamefont {G.}~\bibnamefont
  {Eyink}}\ and\ \bibinfo {author} {\bibfnamefont {H.}~\bibnamefont {Spohn}},\
  }\href@noop {} {\bibfield  {journal} {\bibinfo  {journal} {J. Stat. Phys.}\
  }\textbf {\bibinfo {volume} {70}},\ \bibinfo {pages} {833} (\bibinfo {year}
  {1993})}\BibitemShut {NoStop}%
\bibitem [{\citenamefont {Weiss}\ and\ \citenamefont
  {McWilliams}(1991)}]{wm91}%
  \BibitemOpen
  \bibfield  {author} {\bibinfo {author} {\bibfnamefont {J.}~\bibnamefont
  {Weiss}}\ and\ \bibinfo {author} {\bibfnamefont {J.}~\bibnamefont
  {McWilliams}},\ }\href@noop {} {\bibfield  {journal} {\bibinfo  {journal}
  {Phys. Fluids A}\ }\textbf {\bibinfo {volume} {3}},\ \bibinfo {pages} {835}
  (\bibinfo {year} {1991})}\BibitemShut {NoStop}%
\bibitem [{\citenamefont {Tabeling}(2002)}]{Tabeling2002}%
  \BibitemOpen
  \bibfield  {author} {\bibinfo {author} {\bibfnamefont {P.}~\bibnamefont
  {Tabeling}},\ }\href@noop {} {\bibfield  {journal} {\bibinfo  {journal}
  {Phys. Rep.}\ }\textbf {\bibinfo {volume} {362}},\ \bibinfo {pages} {1}
  (\bibinfo {year} {2002})}\BibitemShut {NoStop}%
\bibitem [{\citenamefont {Zermelo}(1902)}]{zermelo}%
  \BibitemOpen
  \bibfield  {author} {\bibinfo {author} {\bibfnamefont {E.}~\bibnamefont
  {Zermelo}},\ }\href@noop {} {\bibfield  {journal} {\bibinfo  {journal} {Z.
  Math. Phys.}\ }\textbf {\bibinfo {volume} {47}},\ \bibinfo {pages} {201}
  (\bibinfo {year} {1902})}\BibitemShut {NoStop}%
\bibitem [{\citenamefont {Kiessling}\ and\ \citenamefont
  {Wang}(2012)}]{kiessling2012}%
  \BibitemOpen
  \bibfield  {author} {\bibinfo {author} {\bibfnamefont {M.}~\bibnamefont
  {Kiessling}}\ and\ \bibinfo {author} {\bibfnamefont {Y.}~\bibnamefont
  {Wang}},\ }\href@noop {} {\bibfield  {journal} {\bibinfo  {journal} {J. Stat.
  Phys.}\ }\textbf {\bibinfo {volume} {148}},\ \bibinfo {pages} {896} (\bibinfo
  {year} {2012})}\BibitemShut {NoStop}%
\bibitem [{\citenamefont {Esler}\ \emph {et~al.}(2013)\citenamefont {Esler},
  \citenamefont {Ashbee},\ and\ \citenamefont {McDonald}}]{Esler}%
  \BibitemOpen
  \bibfield  {author} {\bibinfo {author} {\bibfnamefont {J.}~\bibnamefont
  {Esler}}, \bibinfo {author} {\bibfnamefont {T.}~\bibnamefont {Ashbee}}, \
  and\ \bibinfo {author} {\bibfnamefont {N.}~\bibnamefont {McDonald}},\
  }\href@noop {} {\bibfield  {journal} {\bibinfo  {journal} {Phys. Rev. E}\
  }\textbf {\bibinfo {volume} {88}},\ \bibinfo {pages} {012109} (\bibinfo
  {year} {2013})}\BibitemShut {NoStop}%
\bibitem [{\citenamefont {Pointin}\ and\ \citenamefont
  {Lundgren}(1976)}]{PointLund}%
  \BibitemOpen
  \bibfield  {author} {\bibinfo {author} {\bibfnamefont {Y.}~\bibnamefont
  {Pointin}}\ and\ \bibinfo {author} {\bibfnamefont {T.}~\bibnamefont
  {Lundgren}},\ }\href@noop {} {\bibfield  {journal} {\bibinfo  {journal}
  {Phys. Fluids}\ }\textbf {\bibinfo {volume} {19}},\ \bibinfo {pages} {1459}
  (\bibinfo {year} {1976})}\BibitemShut {NoStop}%
\bibitem [{\citenamefont {Qi}\ and\ \citenamefont {Marston}(2014)}]{qi14b}%
  \BibitemOpen
  \bibfield  {author} {\bibinfo {author} {\bibfnamefont {W.}~\bibnamefont
  {Qi}}\ and\ \bibinfo {author} {\bibfnamefont {J.~B.}\ \bibnamefont
  {Marston}},\ }\href@noop {} {\bibfield  {journal} {\bibinfo  {journal} {J.
  Stat. Mech.}\ }\textbf {\bibinfo {volume} {2014}},\ \bibinfo {pages} {07020}
  (\bibinfo {year} {2014})}\BibitemShut {NoStop}%
\end{thebibliography}

%

\end{document}